\documentclass[a4paper,12pt]{article}
\DeclareMathSizes{12}{12.5}{10}{10}
\usepackage[left=2.5cm,bottom=3cm,right=2.5cm,top=3cm]{geometry} 

\usepackage{overpic,youngtab}
\usepackage{subfigure}
\usepackage[latin,english]{babel}
\usepackage{amsmath}
\usepackage{amssymb}
\usepackage{epstopdf}
\usepackage{graphics,psfrag,rotating}
\usepackage{mathabx}
\usepackage{graphicx}
\usepackage{dcolumn}
\usepackage{float}
\usepackage{pdflscape}
\usepackage{array}
\usepackage{booktabs}
\usepackage{amscd} 
\usepackage{mathtools}
\usepackage{fancybox}
\usepackage{fix-cm}
\usepackage[colorlinks=true,citecolor=blue,,linktocpage=true,linkcolor=blue,urlcolor=black]{hyperref}
\usepackage{tikz} 
\usetikzlibrary{matrix} 
\usetikzlibrary{positioning} 
\usetikzlibrary{arrows}
\usepackage{titlesec}
\usepackage{abstract}
\usepackage{centernot}

\def\be{\begin{equation}}
\def\ee{\end{equation}}
\def\bea{\begin{eqnarray}}
\def\eea{\end{eqnarray}}
\def\pd{\partial}
\def\a{\alpha}
\def\b{\beta}
\def\g{\gamma}
\def\d{\delta}
\def\m{\mu}
\def\n{\nu}
\def\t{\tau}

\def\l{\lambda}

\def\r{\rho}

\def\s{\sigma}
\def\e{\epsilon}
\def\bi{\begin{itemize}}
	\def\ei{\end{itemize}}

\def\xp{x^\prime}

\definecolor{ogreen}{rgb}{0,0.7,0}

\DeclareGraphicsExtensions{.pdf,.png,.jpg,.gif,.jpeg}
\def\be{\begin{equation}}
\def\ee{\end{equation}}
\def\bea#1\eea{\begin{align}#1\end{align}}
\def\pd{\partial}
\def\a{\alpha}
\def\b{\beta}
\def\g{\gamma}
\def\d{\delta}
\def\e{\epsilon}
\def\k{\kappa}
\def\m{\mu}
\def\n{\nu}
\def\t{\tau}

\def\l{\lambda}

\def\r{\rho}

\def\s{\sigma}
\def\e{\epsilon}
\def\t{\tau}
\def\bi{\begin{itemize}}
	\def\ei{\end{itemize}}

\def\bpm{\begin{pmatrix}}
\def\epm{\end{pmatrix}}

\def\xp{x^{\prime}}

\usepackage{authblk}
\usepackage{setspace}

\begin{document}
	
	\vspace*{-1cm}
\phantom{hep-ph/***} 
{\flushleft
	{{FTUAM-21-xx}}
	\hfill{{ IFT-UAM/CSIC-24-157}}}
\vskip 1.5cm
\begin{center}
	{\LARGE\bfseries  Some aspects of QFT in non-inertial frames.  }\\[3mm]
	\vskip .3cm
	
\end{center}

\vskip 0.5  cm
\begin{center}
	{\large Enrique \'Alvarez and  Jes\'us Anero.}
	\\
	\vskip .7cm
	{
		Departamento de F\'isica Te\'orica and Instituto de F\'{\i}sica Te\'orica, 
		IFT-UAM/CSIC,\\
		Universidad Aut\'onoma de Madrid, Cantoblanco, 28049, Madrid, Spain\\
		\vskip .1cm

		\vskip .5cm
		
		\begin{minipage}[l]{.9\textwidth}
			\begin{center} 
				\textit{E-mail:} 
				\tt{enrique.alvarez@uam.es} and
				\tt{jesusanero@gmail.com} 
			\end{center}
		\end{minipage}
	}
\end{center}
\thispagestyle{empty}

\begin{abstract}
	\noindent
	Some aspects of quantum field theory in a general (i.e. non inertial) frame of Minkowski spacetime are studied. Conditions for the presence of horizons as well as for the modification of  the definition of positive energy solutions are examined. The standard wisdom is confirmed that none of these phenomena is generic. This corroborates that in the general case there is no natural way to define positive frequencies, which is the standard road to define particles. In that sense, (as well as in others)  non inertial frames are similar to curved spacetimes.	
\end{abstract}

\newpage
\tableofcontents
\thispagestyle{empty}
\flushbottom

\newpage
	
\section{Introduction.}
One of the most challenging problems of theoretical physics is  to determine what is the precise relationship between General Relativity and Quantum Mechanics. It seems reasonable to think that in order to do so, a first step would be to clarify what happens in ordinary quantum field theory (QFT) when working in non-inertial frames, but still in the Minkowskian framework. Much work has been done following the pioneering essays of \cite{Fulling}\cite{Unruh} in the particular case when the frame is in constant acceleration with respect to an inertial one. A nice review is \cite{Crispino}. Other interesting references are \cite{Biermann}\cite{UnruhPR}.
\par
  Another topic that awoke much interest is the formulation of QFT in a curved spacetime ({\em confer}  for example the books \cite{Birrell}\cite{Fulling}\cite{Parker}\cite{Wald}). We refrain from giving a complete list of references, which  would have been  exceedingly long. We apologize for it.
\par
The simplest question that comes to mind is to determine what aspects of Unruh's phenomenon are generic, valid for all non-inertial frames, and what are specific to the particular situation of constant acceleration.
\par
The purpose of this paper is to examine some simple aspects of QFT as defined in a generic non-inertial frame. 

Conceptually a frame is a basis of the tangent space $T_x$ at a given point of spacetime $x\in V_4$ .  A family of frames 
\footnote{It should be kept in mind that we are not using the more precise mathematical language in this paper. For example, we should be talking about {\em the moving inertial frame on the Rindler chart}. But we shall stick to the usual physicist's terms.}
will be usually defined at all points of a timelike curve $\g$, which can be thought of as the spacetime trajectory of an observer. We shall denote
\be
\vec{e}_a\equiv e_a^{~\m}\pd_\m
\ee
the latin indices $a=0,\ldots ,3$ label the different vectors, whereas $\m=0,\ldots,3$ are vector indices. Were we in a curved manifold, the former would have been {\em Lorentz indices}, whereas the latter  would be {\em Einstein indices}.
The simplest frames are orthonormal, that is
\be
\vec{e}_a.\vec{e}_b\equiv g_{\m\n}(x)\, e_a\,^\m e_b\,^\n=\eta_{ab}
\ee

A change of coordinates (not necessarily global) in flat space
\be
x^a \rightarrow \xi^\m(x)
\ee
defines a frame through the Jacobian
\be
e^a_{~\m}\equiv {\pd x^a\over \pd \xi^\m}
\ee
but of course not all frames can be interpreted in such a way \cite{Matsuda}; moving frames can be used in curved spaces as well, and this is actually the main tool  Cartan used in his work. 
\par
The local condition for a frame to be equivalent to a change of coordinates in flat space  is that the forms
\be
\underline{e}^a\equiv e^a_{~\m} dx^\m
\ee
are closed
\be
d \underline{e}^a=0
\ee
the fact that the Levi-Civita spin connection is torsionless implies in addition, that
\be
\underline{\omega}^a\,_b \wedge \underline{e}^b=0
\ee

Those are the only frames we are going to be interested at in this paper. At any rate 
the commutator of two vectors in this basis of the tangent space  can also be expressed in this same basis, so that
\be
\left[\vec{e}_a,\vec{e}_b\right]\equiv C_{ab}^c \vec{e}_c
\ee
where $C_{ab}^c $ are the {\em structure constants} of the frame and, as we shall see, uniquely determine the metric connection.  As a matter of fact, the spin connection is given by
\be
\omega_{a|bc}={1\over 2}\left(C_{ac|b}+ C_{ba|c}+ C_{bc|a}\right)
\ee
where
\bea
&\omega_{a|bc}\equiv e_a\,^\m \omega_{\m bc} \nonumber\\
&C_{ac|b}\equiv C_{ac}\,^d\eta_{bd}
\eea
\par

\par
The observer's world line  can be  obtained if desired, from the differential equations
\bea
&{dt\over d\t}= \g\nonumber\\
&{d x^i\over d\t}=\g {dx^i\over dt}\equiv \g v^i\quad i=1,2,3
\eea
In general the three-velocity $v^i$ is not constant, and neither is $\g\equiv (1-v^2)^{- 1/2}$. The function $t(\t)$ is to be obtained by integrating
\be
\t=\int^t {dt^\prime\over \g\left(v\left(t^\prime\right)\right)}
\ee

A standard way of refering observations \cite{Matsuda} made by a hypothetical observer is to define a {\em moving frame}  $\vec{e}_a$ attached to her world line. 

The natural way in our case is to  consider as a  timelike vector the four-velocity itself
\be
e^{a}_{~0}\equiv u^a=\frac{d x^a}{d\t}
\ee
and then define the other three spacelike vectors out of derivatives of the four-velocity, assuming them to be non-vanishing. This procedure does not work for the particular case of {\em inertial observers} for which the first derivative ({\em id est} the {\em acceleration}) vanishes
\be
\dot{u}^a\equiv {d u^a\over d\t}=0
\ee
this means that the frames we are going to study here are generically non-inertial.
\par
A generalization of Frenet-Serret's formulas for Minkowskian curves is found in \cite{Synge} This has been reviewed and exploited in \cite{Letaw}. We assume a world line $x^a(\t)$, with $\t$ the proper time. The first step is  to define the timelike vector
\be
e^{a}_{~0}\equiv u^a
\ee
the normalization implies that
\be  \dot{u}^2+u\ddot{u}=0\ee

The {\em unit first normal} $e^{a}_{~1}$ is defined by
\be
{d e^{a}_{~0}\over d\t}\equiv \dot{u^a}=\kappa_1(x) e^{a}_{~1}
\ee
where $\kappa_1$ is the {\em (first) curvature}
\be
-\kappa_1^2=\dot{u}^2\equiv -a^2\Longrightarrow \kappa_1= a
\ee
and  $a$ is the modulus of the acceleration and we have chosen the first curvature to be positive semidefinite.
\be
e^{a}_{~1}={1\over \kappa_1(x)}\dot{u}^a
\ee
The {\em second normal}, $e^{a}_{~2}$ and the {\em second curvature (torsion)} $\kappa_2$, are now defined through
\be
{d e^{a}_{~1}\over d\t}\equiv \kappa_2(x)e^{a}_{~2}+\kappa_1(x) e^{a}_{~0}
\ee
the normalization factor yields
\be \k_2^2=-\dot{e}_1^2-\dot{e}_0^2\ee
explicitly
\be \k_2^2=-\left[\frac{\ddot{u}}{a}-\frac{\dot{a}\dot{u}}{a^2}-au\right]^2\ee
Finally the {\em third normal} and the {\em third curvature (hypertorsion)} are defined by
\bea
&\frac{d e^{a}_{~2}}{d\t}\equiv\kappa_3(x) e^{a}_{~3}-\kappa_2(x) e^{a}_{~1}\nonumber\\
& \frac{d e^{a}_{~3}}{d\t}\equiv-\kappa_3(x)e^{a}_{~2}
\eea
with
\be \k_3^2=-\dot{e}_2^2+\dot{e}_1^2+\dot{e}_0^2\ee
in fact
\be \dot{e}_3^2-\dot{e}_2^2+\dot{e}_1^2+\dot{e}_0^2=0\ee
the tetrad thus obtained is orthonormal $e_a\cdot e_b=\eta_{ab}$. Finally, when evolving in time, the tetrad rotates
\be
\frac{d e^{a}_{~\m}}{d\t}= \Sigma^a_{~b}(\t) e^{b}_{~\m}
\ee
(with $\Sigma_{ab}=-\Sigma_{ba}$), according to \cite{Synge} any curve in spacetime obeys the generalized Frenet-Serret formulas which boils down to
\be
\Sigma_{~b}^{a}=\left(
\begin{array}{cccc}
	0& \k_1& 0&0\\
	\k_1& 0& \k_2&0\\
	0& -\k_2& 0&\k_3\\
	0&0&-\k_3&0
\end{array}
\right)
\ee
the simplest curves  are  those that have got all coefficients $\k_1,\k_2,\k_3$  constants, and are usually called {\em helixes (stationary motions)}.

\par
Before presenting a list of all stationary motions, it is worth remarking that
\be
\frac{d e_0}{d\t}=\dot{u}=\sum_{a=1}^3 (\dot{u}e_a)e^a=\sum_{a=1}^3 a_a e^a
\ee
because $u\dot{u}=\dot{u}e_0=0$.  It is not possible that $\dot{u}e_a=0$ for all the spacelike $e_a$ because this would mean that $\dot{u}=0$. Physically this means that when $a\neq 0$; that is, for any accelerated observed, it is not possible to choose 
$\Sigma_{ab}=0$. The minimal rotation (usually called Fermi-Walker) is given by
\be
\Sigma_{ab}=\begin{pmatrix}0&a_1&a_2&a_3\\-a_1&0&0&0\\-a_2&0&0&0\\-a_3&0&0&0\end{pmatrix}
\ee
\be
\Sigma_{ab}=\d_{a0}\d_{bi}a_i-\d_{ai}\d_{b0}a_i
\ee

The stationary motions (helixes)  can be classified in six classes
\begin{enumerate}
	\item $\k_1=\k_2=\k_3=0$ with 
	\be e^{a}_{~0}=\left(1, 0, 0,0\right)\ee
	\item $\k_2=\k_3=0$ with 
	\be e^{a}_{~0}=\left(\cosh \k_1\t, \sinh \k_1\t, 0,0\right)\ee
	\item $|\k_1|<|\k_2|$ and $\k_3=0$ with 
	\be e^{a}_{~0}=\frac{1}{\r^2}\left(\k_2^2-\k_1^2\cos \r\t, \k_1\r \sin \r\t, \k_1\k_2(1-\cos\r\t),0\right)\ee 
	where $\r=\k_2^2-\k_1^2$
	\item $|\k_1|=|\k_2|$ and $k_3=0$ with 
	\be e^{a}_{~0}=\left(1+\frac{1}{2}\k_1^2\t^2, \k_1\t ,\frac{1}{2}\k_1^2\t^2,0\right)\ee
	\item$|\k_1|>|\k_2|$ and $\k_3=0$ with 
	\be e^{a}_{~0}=\frac{1}{\s^2}\left(-\k_2^2+\k_1^2\cosh \s\t, \k_1\s \sinh \s\t, -\k_1k_2(1-\cosh\s\t),0\right)\ee 
	where $\s=\k_2^2-\k_1^2$
	\item $\k_3\neq0$ with 
	\be e^{a}_{~0}=A^a\cosh R_{+}\t+B^a\sinh R_{+}\t+C^a\cos R_{-}\t+D^a\sin R_{-}\t\ee 
	where 
	\bea 
	&A^a=\frac{1}{R^2}\left(R_{-}^2+\k_1^2,0,\k_1\k_2,0\right)\nonumber\\
	&B^a=\frac{1}{R^2}\left(0,\k_1(R_{-}^2+\k_1^2-\k_2^2)/R_{+},0,\k_1\k_2\k_3/R_{+}\right)\nonumber\\
	&C^a=\frac{1}{R^2}\left(R_{+}^2-\k_1^2,0,-\k_1\k_2,0\right)\nonumber\\
	&D^a=\frac{1}{R^2}\left(0,\k_1(R_{+}^2-\k_1^2+\k_2^2)R_{-},0,-\k_1\k_2\k_3/R_{-}\right)
	\eea
	with $R^2=R_{-}^2+R_{+}^2$
	\end{enumerate}

\par

The contents of this paper are as follows.

 After a brief recap of Killing vectors and their horizons, we study in some detail a few explicit examples, namely the Rindler frame, the  drifted Rindler frame, an harmonic frame, a rotating frame, a cusped (parator) frame, an interpolating frame, and a general helix. Most of the examples  have non-constant acceleration. We also determine whether the tangent vector is a linear combination of Killing vectors.
 
 Then we employ Wiener-Khinchin's  theorem to determine when there would be a detector's response.  After that, we state our conclusions.
\section{Reminder of Killing vectors.}

There are particular diffeomorphisms that leave invariant the metric. Those are dubbed {\em isometries}. The generator of an isometry is called a Killing field. Whenever they exist, they are specific to a given metric, and obey
\be \pounds(k)g_{\a\b}=\nabla_\a k_\b+\nabla_\b k_\a=k^\l\partial_\l g_{\a\b}+g_{\l\a}\partial_\b k^\l+g_{\l\b}\partial_\a k^\l\ee
it is easy to prove that
\be\label{e}
\nabla_\m\nabla_\r k_\n=R_{\m\n\r\l}k^\l
\ee

Assume all isometries of a given manifold are given by $ \vec{k}_a\equiv k_a^\m\pd_\m$, the commutator of two such isometries can be proved to be  another isometry. 
\be
\left[\vec{k}_a,\vec{k}_b\right]=C_{ab}^c   \vec{k}_c
\ee

The set of all isometries constitute a group, which is {\em simply transitive} if the Killing  vectors are linearly independent. Otherwise  the group is multiply transitive. The orbits of the group are homogeneus spaces, which have the same dimension as the group in the simply transitive case.
When a given space enjoys a timelike isometry the space is {\em stationary}. It is useful to consider {\em coordinates adapted to the isometry}, in which the isometry reads
\be
k^a=(1,0,0,0)
\ee
then
\be
k_0= g_{00}
\ee
there is no need for $g_{0i}=0$. When this is also the case, we say that the spacetime is {\em static}. Ay any rate the equation for the isometry in adapted coordinates guarantees that time is a cyclic coordinate
\be
\pd_0 g_{a b}=0
\ee

All this applies equally to flat space. There are 10 Killing vectors, which can be interpreted as the generators of the Poincar\'e group, which includes rotations, boosts, and spacetime translations.

In cartesian coordinates where the Christoffel symbols vanish they read
\begin{equation}
\begin{array}{l}
{P_a=\partial_a}\\[4pt]
{L_{ab}=x_a\partial_b-x_b\partial_a}
\end{array}
\end{equation}
$P_0$ is timelike in the whole $M_4$, and $L_{0i}$ is timelike in the region $x_i^2-x_0^2 >0$. The rest of the Killings are everywhere spacelike.

The general Killing vector can be written as
\begin{equation}
\begin{array}{l}
{\sum_a \xi^a P_a+\sum_i\l^i L_{0i}+\sum_{ij} \m^{ij}L_{ij}=}\\[4pt]
{=\left(\xi^0-\sum_i \l^i x_i\right)\partial_0+\sum_j\left(\xi^j+x_0\l^j+\sum_k(\m^{kj}-\m^{jk})x_k\right)\partial_j}
\end{array}
\end{equation}
then the Killing reads
\be k^\m\equiv\left(\xi^0-\sum_i \l^i x_i,\xi^j+x_0\l^j+\sum_k(\m^{kj}-\m^{jk})x_k\right)\ee
it will be timelike whenever
\be \left(\xi^0-\sum_i \l^i x_i\right)^2-\left(\xi^j+x_0\l^j+\sum_k(\m^{kj}-\m^{jk})x_k\right)\left(\xi_j+x_0\l_j+\sum_k(\m^{k}_{~j}-\m_j^{~k})x_k\right)>0\ee

The fact that Riemann's tensor vanishes, means that our previous equation (\ref{e})  implies that
\be
\nabla_\m\nabla_\r k_\n=0
\ee
this is not as trivial as it seems, because in non-cartesian coordinates Christoffel's symbols fail to vanish in general.
\par
Given any frame that can be associated to a timelike observer
\be
x^a=x^a(\t),
\ee
it is just natural to assume that the observer's proper time is the natural measure of time for her. The tangent vector
\be
{\pd\over \pd \t}=u^\m\pd_\m
\ee
 is normalized to unity. Nevertheless, in order to be allowed to use the tangent vector to define {\em positive frequencies } (which is the first step in the definition of {\em particles}) it is necessary for  this vector to be  a linear combination of Killing vector fields. This is not always the case. As a matter of fact  both $\g$ and $u^i$ depend in general on $\t$. Only in the case where $\dot{v}^i=0$ we have
\be
{\pd\over \pd \t}= t{\pd\over \pd  t}+x^i {\pd\over \pd x^i}
\ee
which belongs to the Killing algebra. Otherwise this only happens in a few lucky cases. We shall encounter examples of both lucky and unlucky situations in the sequel.
\par
It is well known that the root of Unruh's effect is precisely the fact that what amounts to the {\em natural} definition of positive frequencies (and consequently, upon quantization, particles) in the non-inertial frame is a combination of positive and negative frequencies as defined with respect to the inertial frame. This was first noted in this language by Fulling \cite{Fulling}, although it was foreshadowed by Schr\" odinger \cite{Schrodinger} as early as  1939.
\par
This means that all Green functions that are defined as vacuum expectation values (vev) with some time-ordering (namely, the advanced, retarded, Feynman and Dyson) do {\em not} behave as  true scalars under change of frame because the {\em vacuum changes}. Then
\be
G^\prime(\xp)\neq G(x)
\ee
all quantities referring to field operators (no vev involved) are still {\em bona fide} tensors. This is the case for the energy-momentum tensor and the ensuing four-momentum
\be
P^\m(\Sigma)=\int d\Sigma_\n\, T^{\m\n}
\ee
the energy as measured in some frame characterized by the tangent vector $u$ (assumed orthogonal to the spacelike surface $\Sigma$) is some vev
\be
E(u)\equiv \langle 0 \left|u^\m P_\m\right|0\rangle
\ee
even for fixed $u$, this depends on the vacuum state $|0\rangle$. It could be natural to {\em define} the vacuum state $|0_u\rangle$ as the one minimizing $E(u)$, but this condition is not in general sufficient to uniquely determine it.
\par
Adapted coordinates $\xi^\m(x)$ to a given observer are by definition, such that in them
\be 
u^\m(\xi)\equiv \frac{\partial \xi^\m}{\partial x^a}u^a\equiv e^{~\m}_a(x) u^a(x)=\d^\m_0\ee
this is a set of partial differential equations (PDE)  for the mapping
\be
x^a\in M_4\rightarrow \xi^\m \in M_4
\ee

The inverse Jacobian is
\be
e^a_{~\m}=\frac{\partial x^a}{\partial \xi^\m}
\ee
and the inertial frames are those for which
\be
\eta_{ab}\, e^a_{~\m} e^b_{~\n}=\eta_{\m\n}
\ee
those are generated by isometries of Minkowski space. Otherwise the frame is non inertial, and we define 
\be
g_{\a\b}=\eta_{ab}e^a_{~\a}e^b_{~\b}
\ee

\par
The Killing's norm is of course conserved
\be k^a\frac{\partial}{\partial x^a}=k^a e_a^{~\m}\frac{\partial}{\partial \xi^\m}=k^\m\frac{\partial}{\partial \xi^\m}\ee

\be k^ak^b\eta_{ab}=k^\m k^\n e^a_{~\m}e^b_{~\n}\eta_{ab}=k^\m k^\n g_{\m\n}\ee
in fact, in adapted coordinates
\be
k_a^2(x)=e^a_{~0}(x) e^b_{~0}(x)\eta_{ab}\equiv g_{00}
\ee
horizons are defined by the zeros of the norm of this Killing. Owing to the fact that flat space is static, all horizons are in fact Killing horizons. To be specific, $g_{00}(x)$ is a function
\be
F: x\in V_4\quad \longrightarrow  F(x) \in\mathbb{R}
\ee
the horizons are determined by the {\em zero set} $F^{-1}(0)$. 
\par

The adapted Killing $k_a$ can of course be written in cartesian coordinates, where it got to be a linear combination of all Killing vectors, that is
\be
k_a=\sum_b \xi^b P_b+\sum_i\l^i L_{0i}+\sum_{ij} \m^{ij}L_{ij}
\ee

\section{Some  simple examples with constant and non-constant acceleration.}
Let us first clarify what we understand by constant acceleration. The covariant definition \cite{Rohrlich}
reads
\be
\dot{a}^\m_\perp\equiv \ddot{u}^\m_\perp=\left(\d^\m_\l-u^\m u_\l\right)\ddot{u}^\l=0
\ee
this means that either  
\be
\dot{a}^\m\equiv \ddot{u}^\m=0;
\ee
 or else 
\be
  \ddot{u}= a^2 u
  \ee
  In the next paragraph we shall see an example of this.
\subsection{Rindler frame with constant acceleration.}
The world line of the fidutial observer reads
\be x^a=\frac{1}{a}\Big(\sinh a\t, \cosh a\t, 0,0\Big)\ee

The  orthonormal frame is given by

\bea
&e^{a}_{~0}=\big(\cosh a\t, \sinh a\t, 0,0\big)\nonumber\\
&e^{a}_{~1}=\big(\sinh a\t, \cosh a\t, 0,0\big)\nonumber\\
&e^{a}_{~2}=\Big(0, 0, 1,0\Big)\nonumber\\
&e^{a}_{~3}=\Big(0, 0, 0,1\Big)
\eea
then  $\dot{u}^2=-a^2$, the modulus squared of the acceleration. Besides
\be \ddot{u}^\m_{\perp}=0\ee
because $\ddot{u}^\l=a^2 u^\l$.

The curvatures of the world line are given by
\bea
&\k_1=a\nonumber\\
&\k_2=0\nonumber\\
&\k_3=0
\eea
therefore is Class II in Synge's classification.

Adapted coordinates  are given by
\bea
\tanh\,a\xi^0={t\over x}\quad&\vline\quad t= {1\over a} e^{a\xi^1} \sinh\,a\xi^0\nonumber\\
e^{2 a \xi^1}= a^2(x^2-\t^2)\quad&\vline\quad x= {1\over a} e^{a\xi^1} \cosh\,a\xi^0\nonumber\\
\xi^2=y\quad&\vline\quad y=\xi^2 \nonumber\\
\xi^3=z\quad&\vline\quad z=\xi^3
\eea
whose domain is Rindler's wedge
\be
x^2-t^2\geq 0
\ee
and the range is the full $M_4$. Then, the flat metric in adapted coordinates reads
\be\text{d}s^2=\r^2 \text{d}\omega^2-\text{d}\r^2-\text{d}\xi_2^2-\text{d}\xi_3^2
\ee
where
\bea
\xi^0=\frac{1}{a}\omega\quad&\vline\quad t=\r \sinh\,\omega\nonumber\\
e^{a\xi^1}=a\r\quad&\vline\quad x=\r\cosh\,\omega\nonumber\\
\xi^2=y\quad&\vline\quad y=\xi^2 \nonumber\\
\xi^3=z\quad&\vline\quad z=\xi^3
\eea
it does not seem possible however for adapted coordinates to range over the whole $M_4$ with a single local frame.
\par
\par
The most important thing to notice is that the tangent vector to the trajectory is proportional to a Killing vector, namely $L_{{01}}$, namely
\be \frac{\partial}{\partial \t}=u^a\frac{\partial}{\partial x^a}=a\left(t\frac{\partial}{\partial x}+x\frac{\partial}{\partial t}\right)=a L_{01}\ee
and this is the natural Killing to choose in order to define positive frequencies. As we have already mentioned, and we shall see in detail in a moment, this is exceptional in trajectories with nontrivial acceleration.
\subsection{The interpolating motion}
It should be remarked that in \cite{Parry} it is argued that circular acceleration can be viewed as dual to the drifted Rindler motion, whose Killing vector is a linear combination of a boost and a spatial translation and in fact these motions can be smoothly deformed one into the other by means of the   {\em cusped}  or {\em parator} motion. This interpolation is provided by the frame associated to the world line given by
\be
x^a(\t)=\left(\g^2\t-\g vR\sin\left[ \frac{\g v\t}{R}\right],R\cos \left[ \frac{\g v\t}{R}\right]-R, \g R\sin \left[ \frac{\g v\t}{R}\right]-\g^2 v\t,0\right)
\ee
where $\g=(1-v^2)^{-\frac{1}{2}}$, in this case the acceleration is not constant.

The associated orthonormal frame is
\bea
&e^{a}_{~0}=\left(\g^2\left(1-v^2\cos \left[ \frac{\g v\t}{R}\right]\right), -v\g\sin \left[ \frac{\g v\t}{R}\right], -v\g^2\left(1-\cos \left[ \frac{\g v\t}{R}\right]\right),0\right)\nonumber\\
&e^{a}_{~1}=\left(v\g\sin\left[ \frac{\g v\t}{R}\right], \cos \left[ \frac{\g v\t}{R}\right], -\g\sin \left[ \frac{\g v\t}{R}\right],0\right)\nonumber\\
&e^{a}_{~2}=\left(-v\g^2\left(1-\cos \left[ \frac{\g v\t}{R}\right]\right), \g\sin \left[ \frac{\g v\t}{R}\right], \g^2\left(v^2-\cos\left[ \frac{\g v\t}{R}\right]\right),0\right)\nonumber\\
&e^{a}_{~3}=\left(0, 0, 0,1\right)
\eea

The corresponding curvatures are given by
\bea
&\k_1=\frac{v^2\g^2}{R}\nonumber\\
&\k_2=\frac{v\g^2}{R}\nonumber\\
&\k_3=0
\eea
it is therefore Synge's Class V and $v=\frac{ºk_1}{\k_2}$

The vector tangent to the trajectory is
\be  \frac{\partial}{\partial \t}=\frac{\partial}{\partial t}-k_1\left(x\frac{\partial}{\partial t}+t\frac{\partial}{\partial x}\right)+k_2\left(x\frac{\partial}{\partial y}-y\frac{\partial}{\partial x}\right)=P_0-\k_1L_{01}+\k_2L_{12}\ee
this means that the associated Killing vector is given by
\be k^\m=\left(1-\k_1 x,-\k_1 t-\k_2 y, -\k_2x,0\right)\ee

\subsection{Drifted Rindler motion.}

The  world line corresponding to drifted Rindler motion is given by
\be
x^a(\t)=\left(R\sinh \frac{\g\t}{R},R\cosh \frac{\g\t}{R}, \sqrt{\g^2-1}\t,0\right)
\ee
it is easy to check that the acceleration is not constant.

The orthonormal frame attached to the world line reads
\bea
&e^{a}_{~0}=\left(\g\cosh \frac{\g\t}{R}, \g\sinh \frac{\g\t}{R}, \sqrt{\g^2-1},0\right)\nonumber\\
&e^{a}_{~1}=\left(\sinh \frac{\g\t}{R}, \cosh \frac{\g\t}{R}, 0,0\right)\nonumber\\
&e^{a}_{~2}=\frac{1}{ \sqrt{\g^2-1}}\left((1-\g^2)\sinh \frac{\g\t}{R}, (1-\g^2)\cosh \frac{\g\t}{R}, -\g\sqrt{\g^2-1},0\right)\nonumber\\
&e^{a}_{~3}=\left(0, 0, 0,1\right)
\eea

The corresponding curvatures read
\bea
&\k_1=\frac{\g^2}{R}\nonumber\\
&\k_2=\frac{\g\sqrt{\g^2-1}}{R}\nonumber\\
&\k_3=0
\eea
it is therefore a Synge Class V helix.

The vector tangent to the trajectory is given by
\be  \frac{\partial}{\partial \t}=\frac{\g}{R}\left(x\frac{\partial}{\partial t}+t\frac{\partial}{\partial x}\right)+\sqrt{\g^2-1}\frac{\partial}{\partial y}\ee
and it is indeed a linear combination of flat space Killing vectors.
\par
There is a horizon located at
\be
\g^2(x^2-t^2)=R^2(\g^2-1)
\ee
\subsection{Cusped (parator) motion.}
The world line   reads
\be x^a=\left(\t+\frac{1}{6}a^2\t^3, \frac{1}{2}a\t^2, \frac{1}{6}a^2\t^3,0\right)\ee
also here the acceleration fails to be constant.
The attached orthonormal frame is
\bea
&e^{a}_{~0}=\left(1+\frac{1}{2}a^2\t^2, a\t, \frac{1}{2}a^2\t^2,0\right)\nonumber\\
&e^{a}_{~0}=\left(a\t,1, a\t,0\right)\nonumber\\
&e^{a}_{~0}=\left(-\frac{1}{2}a^2\t^2, -a\t, 1-\frac{1}{2}a^2\t^2,0\right)\nonumber\\
&e^{a}_{~0}=\left(0, 0, 0,1\right)
\eea

Curvatures are given by
\bea
&\k_1=a\nonumber\\
&\k_2=a\nonumber\\
&\k_3=0
\eea
we are therefore dealing with a Synge's Class IV helix.

The vector tangent to the trajectory is
\be  \frac{\partial}{\partial \t}=(1+a y)\left(\frac{\partial}{\partial t}+\frac{\partial}{\partial y}\right)+a(t-z)\frac{\partial}{\partial x}=  \frac{\partial}{\partial \t}=(1+a y)\left(\frac{\partial}{\partial t}+\frac{\partial}{\partial y}\right)+\sqrt{2ay}\frac{\partial}{\partial x}\ee
it fails to be a linear combination of Killing vectors.
\subsection{Harmonic trajectory.}
Consider next a {\em harmonic  frame}  defined  along the following simple one-dimensional harmonic trajectory
\bea
&x= {\cos\,\omega t\over \omega}\nonumber\\
&y=0\nonumber\\
&z=0
\eea
with constant frequency $\omega$. The corresponding world line is given by
\be
x^a(\t)={1\over \omega}\left(\arcsin\, \omega\t,\sqrt{1-\omega^2\t^2}-1,0,0\right)
\ee
where
\be
\sin \omega t=\omega \t
\ee
the orthonormal frame is
\bea
&e^{a}_{~0}=\cfrac{1}{\sqrt{1-\omega^2 \t^2}}\big(1, -\omega \t, 0,0\big)\nonumber\\
&e^{a}_{~1}=\cfrac{1}{\sqrt{1-\omega^2 \t^2}}\big(\omega \t, -1, 0,0\big)\nonumber\\
&e^{a}_{~2}=\Big(0, 0, 1,0\Big)\nonumber\\
&e^{a}_{~3}=\Big(0, 0, 0,1\Big)
\eea

The curvatures are given by
\bea
&\k_1={\omega\over \sqrt{1-\omega^2\t^2}}\nonumber\\
&\k_2=0\nonumber\\
&\k_3=0
\eea
it then belongs to Synge's  Class II.

This frame represent a particle with non-constant acceleration
\be a={\omega\over \sqrt{1-\omega^2 \t^2}}\ee
in fact, $\ddot{u}^\l\neq a^2 u^\l$, and the modulus of the acceleration diverges at the boundary where 
\be
\omega t=\pm {\pi\over 2}
\ee
\par
The adapted frame is defined by
\bea
\xi^0=x+\cfrac{1}{\omega}\big(\sin\omega t-\cos\omega t\big)\quad&\vline\quad t=\cfrac{1}{\omega}\arcsin[\omega(\xi^0-\xi^1)]\nonumber\\
\xi^1=x-\cfrac{1}{\omega}\cos\omega t\quad&\vline\quad x=\xi^1+\cfrac{1}{\omega}\sqrt{1-\omega^2(\xi^0-\xi^1)^2}\nonumber\\
\xi^2=y\quad&\vline\quad y=\xi^2 \nonumber\\
\xi^3=z\quad&\vline\quad z=\xi^3
\eea
the adapted system is defined only for
\be
(t^\prime-x^\prime)^2\leq {1\over \omega^2}
\ee
which is a two-sheeted hyperbola with vertexes $\pm{1\over \omega}$  in the I-III regions of Rindler's plane.
\par
The metric in the adapted frame reads
\be 
\text{d}s^2=\text{d}t^2-2 \left(1-\tan\,\omega t\right) \text{d}t \text{d}x -2 \tan\,\omega t\, \text{d}x^2-\text{d}y^2-\text{d}z^2
\ee
the fact that in this coordinates $g_{00}=1$ implies that there are no Killing horizons in this case.
The vector tangent to the trajectory is
\be  \frac{\partial}{\partial \t}=\frac{1}{\omega x}\left(\frac{\partial}{\partial t}-\sqrt{1-\omega^2x^2}\frac{\partial}{\partial x}\right)\ee
which is {\em not} proportional to a Killing vector. This means that there is no natural definition of positive frequencies other than the standard one using ${\pd\over \pd t}$.
\subsection{A rotating frame.}
Another interesting example is the {\em rotating quantum vacuum} of \cite{Davies}, which corresponds to a rigid rotation on the plane $z=z_0$. That is
\bea
&z=z_0\nonumber\\
&x=r_0\cos\,\omega\,t\nonumber\\
&y=r_0\sin\,\omega\,t
\eea

The world line  here reads
\be x^a=\frac{1}{\sqrt{1-\omega^2 r_0^2}}\Big(\t,r_0\cos\omega\t, r_0\sin\omega \t, 0\Big)\ee
and the attached orthonormal frame is
\bea
&e^{a}_{~0}=\frac{1}{\sqrt{1-\omega^2 r_0^2}}\Big(1, -\omega  r_0\sin\omega\t,\omega  r_0  \cos\omega\t,0\Big)\nonumber\\
&e^{a}_{~1}=\big(0,\cos\omega \t,\sin\omega\t,0\big)\nonumber\\
&e^{a}_{~2}=-\frac{1}{\sqrt{1-\omega^2 r_0^2}}\Big(\omega r_0,-\sin\omega\t, \cos\omega\t,0\Big)\nonumber\\
&e^{a}_{~3}=\Big(0, 0, 0,1\Big)
\eea

Curvatures are given by
\bea
&\k_1=\frac{-\omega^2 r_0}{\sqrt{1-\omega^2 r_0^2}}\nonumber\\
&\k_2=\frac{\omega }{\sqrt{1-\omega^2 r_0^2}}\nonumber\\
&\k_3=0
\eea
therefore if $|\omega r_0|>1$ it belongs to Class V but if $|\omega r_0|<1$ it belongs to Class III instead.

This frame represent a particle with constant modulus of the acceleration
\be a=\frac{-\omega^2 r_0}{\sqrt{1-\omega^2 r_0^2}}\ee
but the acceleration itself is not constant in the covariant sense, because
\be
\ddot{u}^\m_\perp\neq 0
\ee
as
\be
 \ddot{u}^\l\neq a^2 u^\l
 \ee

The adapted frame is defined by
\bea
\xi^0=t\sqrt{1-\omega^2 r_0^2}\quad&\vline\quad t=\frac{1}{\sqrt{1-\omega^2 r_0^2}}\xi^0\nonumber\\
\xi^1=x+r_0\cos\omega t\quad&\vline\quad x=\xi^1-r_0\cos\left[\frac{\omega\xi^0}{\sqrt{1-\omega^2 r_0^2}}\right]\nonumber\\
\xi^2=y+r_0\sin\omega t\quad&\vline\quad y=\xi^2-r_0\sin\left[\frac{\omega\xi^0}{\sqrt{1-\omega^2 r_0^2}}\right]\nonumber\\
\xi^3=z\quad&\vline\quad z=\xi^3
\eea

The metric in the adapted frame reads
\bea &\text{d}s^2=\text{d}\xi_0^2-\text{d}\xi_1^2-\text{d}\xi_2^2-\text{d}\xi_3^2 +\nonumber\\
&+\frac{2\omega r_0}{\sqrt{1-\omega^2 r_0^2}}\left(\sin\left[\frac{\omega\xi^0}{\sqrt{1-\omega^2 r_0^2}}\right]\text{d}\xi_0\text{d}\xi_1-\cos\left[\frac{\omega\xi^0}{\sqrt{1-\omega^2 r_0^2}}\right]\text{d}\xi_0\text{d}\xi_2\right)\eea
this metric is in fact Riemann flat. Again, the fact that in this coordinates $g_{00}=1$ implies that there are no Killing horizons  in this case.

\par
The vector tangent to the trajectory is
\be \frac{\partial}{\partial \t}=\frac{1}{\sqrt{1-\omega^2 r_0^2}}\frac{\partial}{\partial t}+\omega\left( x\frac{\partial}{\partial y}-y\frac{\partial}{\partial x}\right)\ee
this is indeed a sum of two Killing vectors, namely
\be k^\m=\left(\frac{1}{\sqrt{1-\omega^2 r_0^2}},-\omega y,\omega x,0\right)\ee
and 
\be k_\m k^\m=\frac{1}{1-\omega^2 r_0^2}-\omega^2 r_0^2\ee
which is timelike in the whole region on which the system is well defined, namely 
\be
\frac{1}{1-\omega^2 r_0^2}>\omega^2 r_0^2
\ee
i.e
\be |\omega r_0|<1\ee
otherwise there is a contradiction with special relativity. 
\par
This was thought to be a fatal flaw in \cite{Davies}, which then consider a modified model with finite spatial extent. We rather think that this problem is related to the well-known absence of rigid bodies is special relativity \cite{Pauli} (see nevertheless the analysis in a Friedmann-Robertson-Walker spacetime \cite{AlvarezBel}) but does not generate any inconsistency.
\subsection{A general helix.}
It is well known that a charged particle moves in a constant electromagnetic field in such a way that its world line is always a helix. The most general case with $\vec{E}\cdot\vec{H}\neq 0$ is
\be x^a=\Big(\frac{\sqrt{1+q^2}}{\chi}\sinh[\chi \t],\frac{q}{\omega}\sin[\omega\t], -\frac{q}{\omega}\cos[\omega \t], \frac{\sqrt{1+q^2}}{\chi}\cosh[\chi \t]\Big)\ee
here again the acceleration is not constant. The attached orthonormal frame reads
\bea
&e^{a}_{~0}=\Big(\sqrt{1+q^2}\cosh[\chi \t],q\cos[\omega\t], q\sin[\omega \t], \sqrt{1+q^2}\sinh[\chi \t]\Big)\nonumber\\
&e^{a}_{~1}=\frac{1}{\sqrt{(1+q^2)\chi^2+q^2\omega^2}}\Big(\sqrt{1+q^2}\chi\sinh[\chi \t],-q\omega\sin[\omega\t], q\omega\cos[\omega \t], \sqrt{1+q^2}\chi\cosh[\chi \t]\Big)\nonumber\\
&e^{a}_{~2}=\Big(-q\cosh[\chi \t],- \sqrt{1+q^2}\cos[\omega\t], - \sqrt{1+q^2}\sin[\omega \t],-q\sinh[\chi \t]\Big)\nonumber\\
&e^{a}_{~3}=\frac{1}{\sqrt{(1+q^2)\chi^2+q^2\omega^2}}\Big(-q\omega \sinh[\chi\t],-\sqrt{1+q^2}\chi\sin[\omega\t],\sqrt{1+q^2}\chi\cos[\omega\t],-q\omega \cosh[\chi\t]\Big)
\eea

Curvatures are given by
\bea
&\k_1=\sqrt{(1+q^2)\chi^2+q^2\omega^2}\nonumber\\
&\k_2=\frac{q\sqrt{1+q^2}(\chi^2+\omega^2)}{\sqrt{(1+q^2)\chi^2+q^2\omega^2}}\nonumber\\
&\k_3=-\frac{\chi\omega}{\sqrt{(1+q^2)\chi^2+q^2\omega^2}}
\eea
we are then dealing with a Synge's Class VI

Adapted coordinates $\xi^\m(x)$  read
\bea
&\tanh[\chi\xi^0]=\cfrac{t}{z}\nonumber\\
&e^{2\chi\xi^1}=\chi^2(z^2-t^2)\nonumber\\
&\xi^2=y\sin\left[\frac{\omega\log[t+z]}{\chi}\right]+x\cos\left[\frac{\omega\log[t+z]}{\chi}\right]\nonumber\\
&\xi^3=x\sin\left[\frac{\omega\log[t+z]}{\chi}\right]-y\cos\left[\frac{\omega\log[t+z]}{\chi}\right]\eea
therefore
\bea
&t=\frac{e^{\chi\xi^1}}{\chi}\sinh\chi\xi^0\nonumber\\
&x=\xi^2\cos\phi[\xi^0,\xi^1]+\xi^3\sin\phi[\xi^0,\xi^1]\nonumber\\
&y=\xi^2\sin\phi[\xi^0,\xi^1]-\xi^3\cos\phi[\xi^0,\xi^1]\nonumber\\
&z=\frac{e^{\chi\xi^1}}{\chi}\cosh\chi\xi^0
\eea
where
\be \phi[\xi^0,\xi^1]\equiv\frac{\omega}{\chi}\log\left[\frac{e^{\chi\xi^1}}{\chi}(\sinh\chi\xi^0+\cosh\chi\xi^0)\right]\ee
The adapted metric yields
\bea &\text{d}s^2=\left(e^{2\xi_1\chi}  -  \omega^2(\xi_2^2 + \xi_3^2) \right)\text{d}\xi_0^2-\left( e^{2\xi_1\chi}  +  \omega^2(\xi_2^2 + \xi_3^2)\right)\text{d}\xi_1^2-\text{d}\xi_2^2-\text{d}\xi_3^2-\nonumber\\
&-2 \omega^2(\xi_2^2 + \xi_3^2)\text{d}\xi_0\text{d}\xi_1+2\omega \xi_2(\text{d}\xi_0+\text{d}\xi_1)\text{d}\xi_3-2\omega \xi_3(\text{d}\xi_0+\text{d}\xi_1)\text{d}\xi_2\eea

The vector tangent to the trajectory is
\be \frac{\partial}{\partial \t}=\chi z\cfrac{\partial}{\partial t}-\omega y\cfrac{\partial}{\partial x}+\omega x\cfrac{\partial}{\partial y}+\chi t\cfrac{\partial}{\partial z}\ee
this is indeed a sum of two Killing vectors, namely
\be k^\m=(z\chi,-\omega y,\omega x,t\chi)\ee
and 
\be k_\m k^\m=\chi^2(z^2-t^2)-\omega^2(x^2+y^2)\ee
is timelike in the region
\be
\omega^2(x^2+y^2) < \chi^2(z^2-t^2)
\ee
which marks the presence of a horizon.

\section{Wiener-Khinchin  theorem and Wightmann functions.}
On general grounds it can be argued that   the {\em response function} \cite{Sciama} of a simplified particle detector (cf.\cite{Birrell}\cite{Sciama}  for a discussion) is proportional to 
\bea
{\cal T}(E)&=\int d\t d\t^\prime e^{-iE(\t-\t^\prime)} \langle 0|\phi(x(\t))\phi(x(\t^\prime))|0\rangle=\nonumber\\
&\equiv \int d\t d\t^\prime e^{-iE(\t-\t^\prime)}G^+\left(\phi(x(\t))\phi(x(\t^\prime))\right)
\eea
In fact the Wiener-Khinchin theorem \footnote{Known to Einstein much earlier \cite{Einstein}.} asserts that the energy absorption rate (the {\em power spectrum}) is determined by the Fourier transform of the field autocorrelation function (which is assumed to be   Wightman 's function in our case).
For a free massless scalar the positive frequency Wightman function is given by
\be
G^+\left(x-\xp\right)=-{1\over 4\pi^2}{1\over (t-t^\prime-i\e)^2-\left(\vec{x}-\vec{x}^\prime\right)^2}
\ee
Furthermore, for stationary motions (helixes) Synge's world function (one half the square of the geodesic distance)  depends only on the difference of proper times, $\Omega(\t,\t')=\Omega(\t-\t^\prime)$. 
\par
Then it can be written
\bea
&{\cal T}(E)=\int d\t  e^{-i\,E\t}\,G^+\left(\t\right)
\eea
It is also a fact that for stationary motions at  small proper time
\be
x^a=x^a(0)+u^a \t+O(\t^2)
\ee
so that, redefining the trajectory  in such a way that $x^a(0)=\left(0,0,0,0\right)$
\be
2\Omega(\t)=  \t^2+O(\t^3)
\ee

In the reference \cite{Parry} an astute way to treat the response function was devised.  Start with the Laurent expansion
\be
{1\over  2\Omega(\t)}={1\over (\t-i\e)^2}+ f(\t)
\ee
where $f(\t)$ is analytic in $\t$. Small modifications are in order when one of the components of the world line  is such that ${x(\t)\over \t}$ vanishes when $\t\sim 0$.
This means that we can write
\be
{1\over 2\Omega(\t-i\e)}={1\over 2\Omega(\t-i\e)}-{1\over  (\t-i\e)^2}+{1\over   (\t-i\e)^2}
\ee
The sum of the two first terms is now $f(\t)$ which is analytic, so that the limit when $\e\rightarrow 0^+$ can be taken in the integrand. The third term can be integrated directly.
\be\label{astute}
{\cal T}(E)=-{1\over 4\pi^2}\int_{-\infty}^\infty d\t\,e^{- i E \t}\left({1\over 2\Omega(\t)}-{1\over \t^2}\right)-{E\over 2\pi}\theta(-E)
\ee
since
\be
\int_{-\infty}^\infty  d\t e^{-i\t E}{1\over (\t-i\e)^2}=2\pi E\theta(-E)
\ee
and where the integrand  is free of singularities, and vanishes both when $\t\rightarrow 0$ as well as when $\t\rightarrow \infty$. The response function  is the Fourier transform of a function which is absolutely integrable, so that by the Riemann-Lebesgue theorem \cite{Lighthill} its transform goes to zero as $E\rightarrow \infty$.
\begin{enumerate}
\item
First of all consider an {\bf inertial motion}
\be
\vec{x}=\vec{x_0}+ \vec{v} t=\vec{x_0}+ \vec{v} \g \t
\ee
then
\be
G^+(\t-\t^\prime)=-{1\over 4\pi^2}{1\over (\t-\t^\prime-i\e)^2}
\ee
our previous equation [\ref{astute}] yields immediatly no contribution. This can also be seen because the integral over proper time has to be performed for negative imaginary proper times
\be
Im\,\left(\t-\t^\prime\right) < 0
\ee
whereas the poles are located at 
\be
\left(\t-\t^\prime\right)=i\e.
\ee
\item
Let us now turn our attention to a {\em hyperbolic trajectory} which physically corresponds to constant  acceleration
\be x^a=\frac{1}{a}\Big(\sinh a\t, \cosh a\t, 0,0\Big)\ee
the Wightman function
\be (t-t'-i\e)^2-(\vec{x}-\vec{x}')^2=-\frac{2}{a^2}\left(1-\cosh[a(\t-\t'-i\e)]\right)\ee

We can write
\be G^{+}(\t-\t')=-\frac{a^2}{16\pi^2}\frac{1}{\sinh^2\left[\frac{a(\t-\t'-i\e)}{2}\right]}\ee
and using the expansion \cite{G}
\be \frac{1}{\sin^2a}=\sum_{n\in\mathbb{Z}}\frac{1}{(a-n\pi)^2}\ee
and the fact that  $\sinh(x)=-i\sin(ix)$, the response function reads
\bea \label{acc}\mathcal{T}(E)&=\int_{-\infty}^\infty d\t e^{-iE\t}G^{+}(\t)=-\frac{1}{2\pi}E\sum_{n=1}^{\infty}e^{-\frac{2\pi nE}{a}}=-\frac{1}{2\pi}\frac{E}{e^{\frac{2\pi E}{a}-1}}\eea

This means that the response function has got a thermal spectrum with temperature
\be T=\frac{a}{2\pi}\ee
it is important, however, to stress, as has been done in the standard book \cite{Birrell} that the accelerated detector still would mesure a vanishing expectation value for the energy density
\be
\langle 0\left|T_{\m\n}\right|0\rangle=0
\ee

Next, we will present our examples.
\subsection{The interpolating frame.}
Remember that the corresponding world line is given by
\be
x^a(\t)=\left(\g^2\t-\g vR\sin\left[ \frac{\g v\t}{R}\right],R\cos \left[ \frac{\g v\t}{R}\right]-R, \g R\sin \left[ \frac{\g v\t}{R}\right]-\g^2 v\t,0\right)
\ee
where $\g=(1-v^2)^{-\frac{1}{2}}$.

This means that   the world function is given by
\be2\Omega(\t,\t')=\g^2(\t-\t'-i\e)^2-4R^2\sin^2\frac{\g v(\t-\t'-i\e)}{2R}\ee
then the response function reads
\bea
{\cal T}(E)&=-\frac{1}{4\pi^2}\int_{-\infty}^{\infty} d\t  e^{-iE\t}\left[\frac{1}{\g^2\t^2-4R^2\sin^2\frac{\g v\t}{2R}}-\frac{1}{\t^2}\right]-\frac{E}{2\pi}\theta(-E)
\eea

\subsection{Drifted Rindler motion.}
The corresponding world line is given by
\be
x^a=\left(R\sinh \frac{\g\t}{R},R\cosh \frac{\g\t}{R}, \sqrt{\g^2-1}\t,0\right)
\ee
then the world function in terms of the difference of proper times reads
\be2\Omega(\t,\t')=4R^2\sinh^2\frac{\g(\t-\t'-i\e)}{2R}-(\g^2-1)(\t-\t'-i\e)^2\ee
then the response function is given by
\bea
{\cal T}(E)&=-\frac{1}{4\pi^2}\int_{-\infty}^{\infty} d\t  e^{-iE\t}\left[\frac{1}{4R^2\sinh^2\frac{\g\t}{2R}-(\g^2-1)\t^2}-\frac{1}{\t^2}\right]-\frac{E}{2\pi}\theta(-E)
\eea

\subsection{Harmonic motion.}
The harmonic frame is associated to the world line
\be x^a=\frac{1}{\omega}\Big(\arcsin \omega\t, \sqrt{1-\omega^2\t^2}, 0,0\Big)\ee
where  Synge's world function is
\be 2\Omega(\t,\t')=\frac{1}{\omega^2}\left[\left(\arcsin \omega\t-\arcsin \omega\t'\right)^2-\left(\sqrt{1-\omega^2\t^2}- \sqrt{1-\omega^2\t'^2}\right)^2\right]\ee
the response function will be given by
\bea
{\cal T}(E)&=-\frac{1}{4\pi^2}\int_{-\infty}^{\infty} d\t  e^{-iE\t}\left[\frac{\omega^2}{\arcsin^2 \omega\t-(\sqrt{1-\omega^2\t^2}-1)^2}-\frac{1}{\t^2}\right]-\frac{E}{2\pi}\theta(-E)
\eea
\subsection{The rotating frame.}
The world line supporting the frame is 
\be x^a=\g\Big(\t,r_0\cos \omega\t, r_0\sin \omega \t, 0\Big)\ee
the corresponding world function  in terms of the difference of proper times reads
\be2\Omega(\t,\t')=\frac{1}{1-r_0^2\omega^2}\left[(\t-\t'-i\e)^2-4r_0^2\sin^2\frac{\omega(\t-\t'-i\e)}{2}\right]\ee
then the response function
\bea
{\cal T}(E)&=-\frac{1}{4\pi^2}\int_{-\infty}^{\infty} d\t  e^{-iE\t}\left[\frac{1-r_0^2\omega^2}{\t^2-4r_0^2\sin^2\frac{\omega\t}{2}}-\frac{1}{\t^2}\right]-\frac{E}{2\pi}\theta(-E)
\eea
\subsection{Cusped (parator) motion.}
The world line  here reads
\be x^a=\left(\t+\frac{1}{6}a^2\t^3, \frac{1}{2}a\t^2, \frac{1}{6}a^2\t^3,0\right)\ee
then the associated world function reads
\bea2\Omega(\t,\t')&=\left(\t-\t'+\frac{1}{6}a^2(\t^3-\t'^3)-i\e\right)^2-\frac{1}{4}a^2\left(\t^2-\t'^2-i\e\right)-\nonumber\\
&-\frac{1}{36}a^4\left(\t^3-\t'^3-i\e\right)^2\eea
and the response function
\bea
{\cal T}(E)&=-\frac{1}{4\pi^2}\int_{-\infty}^{\infty} d\t  e^{-iE\t}\left[\frac{1}{\t^2+\frac{a^2}{12}\t^4}-\frac{1}{\t^2}\right]-\frac{E}{2\pi}\theta(-E)=\nonumber\\
&=\frac{a}{16\sqrt{6\pi^3}}e^{-\frac{2\sqrt{3}E}{a}}-\frac{E}{2\pi}\theta(-E)
\eea
This response can be argued to be similar to the high energy limit of a thermal function.
\subsection{The general helix.}
The world line of the helix reads
\be x^a=\Big(\frac{\sqrt{1+q^2}}{\chi}\sinh[\chi \t],\frac{q}{\omega}\sin[\omega\t], -\frac{q}{\omega}\cos[\omega \t], \frac{\sqrt{1+q^2}}{\chi}\cosh[\chi \t]\Big)\ee
the corresponding world function is
\be2\Omega(\t,\t')=\frac{4(1+q^2)}{\chi^2}\sinh^2\left[\frac{\chi(\t-\t'-i\e)}{2}\right]-\frac{4q^2}{\omega^2}\sin^2\left[\frac{\omega(\t-\t'-i\e)}{2}\right]\ee
and the response function is given by
\bea
{\cal T}(E)&=-\frac{1}{4\pi^2}\int_{-\infty}^{\infty} d\t  e^{-iE\t}\left[\frac{1}{\frac{4(1+q^2)}{\chi^2}\sinh^2\left[\frac{\chi\t}{2}\right]-\frac{4q^2}{\omega^2}\sin^2\left[\frac{\omega\t}{2}\right]}-\frac{1}{\t^2}\right]-\frac{E}{2\pi}\theta(-E)
\eea

\subsection{Summary}

To summarize, all the stationary motion, including  the Rindler motion which acts as the template, have got a world function whose  expansion around $\t=0$ reads
\be\frac{1}{2\Omega(\t)}=\frac{1}{\t^2}+\mathcal{O}(\t)\ee
On the other hand the response function can be expanded at high or low energies using the expansion of the integrand at low or high proper time, respectively
\be
{\cal T}(E)=-\frac{1}{4\pi^2}\int_{-\infty}^{\infty} d\t  e^{-iE\t}{1\over 2 \Omega(\t)}=-{1\over 4\pi^2 E}\int_{-\infty}^{\infty}  dt e^{-i t}{1\over 2 \Omega({t\over E})}
\ee
with the trick of substracting and adding the most divergent piece,
\bea
{\cal T}(E)&=-\frac{1}{4\pi^2}\int_{-\infty}^{\infty} d\t  e^{-iE\t}\left[\frac{1}{2\Omega(\t)}-\frac{1}{\t^2}\right]-\frac{E}{2\pi}\theta(-E)
\eea
at high energies (which correspond to  $\t=0$ in the integrand)
\be
{\cal T}(E)=\frac{1}{2\pi}\frac{\k_1^2}{12}\d(E)-\frac{E}{2\pi}\theta(-E)
\ee
where $\k_1$ is the first curvature.

On the other hand, at low energies ($\t\rightarrow\pm\infty$ in the integrand) the response function reduces to
\be
{\cal T}(E)=-\frac{E}{2\pi}\theta(-E)
\ee
\end{enumerate}
\section{Conclusions.}

In this paper we have examined some of the distinct phenomena that occur if QFT in generic (non inertial) frames. They are as follows. 
\par
First of all, it usually happens that adapted coordinates  to the world line of a generic observer do not cover the whole Minkowski space, and a horizon of some sort appears in the boundary. Given the fact that the adapted metric is still flat, which means that it is also static, this is necessarily a Killing horizon.
\par
The only Killing that is timelike in the whole Minkowski space is ${\pd\over \pd t}$.  Boosts $L_{0 i}$ are timelike in the region $x_i^2-t^2 > 0$ whose boundary is a Killing horizon for them. Appearance of a horizon is then related to the  advancement  of some particular boost as defining the {\em natural} time direction. It may happen that the would be horizon is outside the domain in which the motion is well defined. This happens for our example of the rotating motion.
\par
Independently of that, it may or may not happen that the {\em natural} positive frequency basis in the generic frame is a combination of positive and negative solutions in cartesian Minkowski coordinates. When this is so, the two basis are related  by a Bogoliubov trasformation, as was first pointed out by  Fulling  \cite{Fulling}.
\par

 We have studied in detail examples of several types of behavior. Except for the template (Rindler motion) all our examples have got non constant acceleration, in spite of the fact that all are stationary motions (constant curvatures).
 
 The fact that  in many cases (in the examples we have studied this happens for both the cusped and the harmonic motions) the tangent vector is {\em not} a linear combination of Killing vectors means that it is in general not possible   the use of ${\pd\over \pd \t}$ to define positive frequencies.
 \par
 The response function  is not similar to Planck's for any of our examples (except for the template) and may be the cusped motion at high temperature.  Nevertheless, it has been proposed  \cite{Good} that sometimes it is possible to define an {\em effective temperature} that captures some essential aspects of the physical situation.

 \par
 In fact the more general treatment of QFT in nontrivial curved spacetimes \cite{Witten} backs up  all the points made in this paper, there is no preferred vacuum state and no natural definition of positive and negative frequencies, so that the concept of particle is not a natural one. What we have corroborated here is that {\em even in flat space} all these restrictions hold. It then seems that the concept of {\em particle} is not only a Minkowskian one, but also  it is restricted to inertial or otherwise quite special observers. 
 \par
 An interesting fact that has been recently reported  is that there appear some novel physical effects (namely, a softening) of the QCD phase transition in accelerated systems \cite{Chernodub}. This lies however outside the scope of our present work.
Let us also comment that although it has been claimed \cite{Witten} that the algebraic (Haag's \cite{Haag}) approach to QFT is more appropiate to deal with certain issues derived from diffeomorphism invariance, we have decided to follow in this paper a very conventional Fock space low brow approach to QFT for the sake of simplicity.

 \par
Experimental verification of these ideas is important, as for any scientific prediction. This is not easy because even in the simplest Unruh's phenomenon, the linear
acceleration needed to reach a temperature of one degree  Kelvin  is of order $2.47\times 10^{20} m/s^2$ \cite{Crispino}. Some proposals using charges in circular accelerators have been put forward by Bell and Leinaas  (quantum fluctuation  of electron orbits in ideal storage rings) \cite{Bell} and by Cozella et al \cite{Cozzella} using classical electrodynamics. In spite of all this impressive work,  our 
feeling  is that there is still no clean and convincing experimental check  even in this case.

\section{Acknowledgements.}

One of us (EA) is grateful for  useful correspondence with Jos\'e Senovilla and Enric Verdaguer.  We acknowledge partial financial support by the Spanish MINECO through the Centro de excelencia Severo Ochoa Program  under Grant CEX2020-001007-S  funded by MCIN/AEI/10.13039/501100011033.
We also acknowledge partial financial support by the Spanish Research Agency (Agencia Estatal de Investigaci\'on) through the grant PID2022-137127NB-I00 funded by MCIN/AEI/10.13039/501100011033/ FEDER, UE.
All authors acknowledge the European Union's Horizon 2020 research and innovation programme under the Marie Sklodowska-Curie grant agreement No 860881-HIDDeN and also byGrant PID2019-108892RB-I00 funded by MCIN/AEI/ 10.13039/501100011033 and by ``ERDF A way of making Europe''.


\hrulefill


\end{document}